\documentclass[12pt]{article}%
\usepackage{amsmath}
\usepackage{amsfonts}
\usepackage{amssymb}
\usepackage{graphicx}%
\providecommand{\U}[1]{\protect\rule{.1in}{.1in}}

\begin{document}

\title{Hyperfine anomalies in Gd and Nd}

  \author{J.R. Persson\\
  NTNU\\
  NO-7491 Trondheim\\
  Norway\\
  E-mail: jonas.persson@ntnu.no}



\maketitle

\begin{abstract}
The hyperfine anomalies in Gd and Nd have been extracted from the experimental
hyperfine structure constants using a new method. In addition to the values of
the hyperfine anomaly new improved values of the nuclear magnetic dipole
moment ratios are derived.

\end{abstract}


\section{Introduction}

The study of hyperfine structure (hfs) in atoms has provided information of
the electromagnetic moments of the nucleus as well as information of the
properties of the electrons in the atom \cite{otten,lindgrenrosen}. The
magnetic hfs has in addition proved to give information on the distribution of
magnetization in the nucleus through the so called Bohr-Weisskopf effect
(BW-effect) \cite{bohrweisskopf,Fujitaarima,Buttgenbach}. Bohr and Weisskopf
\cite{bohrweisskopf} showed that the magnetic dipole hyperfine interaction
constant ($a$-constant) is smaller for an extended nucleus compared with a
point nucleus. The extended charge distribution of the nucleus gives rise to
the so-called Breit-Rosenthal effect
(BR-effect)\cite{Breit,Crawford,Pallas,Rosenberg}. It was also shown that
isotopic variations, in combination with the different contributions to the
hfs from the orbital and spin parts of the magnetization in an extended
nucleus, could yield large isotopic deviations from the point nucleus. The
reason for this is that s- and $p_{1/2}$-electrons have a probability of being
inside the nucleus, thus probing the isotopic change in charge distribution as
well as the distribution of magnetization. In this case, as in most cases, the
differential BR-effect is negligible when two isotopes are compared. The
BR-effect is therefore neglected in the following discussion. The differential
hyperfine anomaly $^{1}\Delta^{2}$, the difference of the BW-effect between
two isotopes, is normally defined as:
\begin{equation}
1+{^{1}\Delta^{2}}={{\frac{a^{(1)}}{a^{(2)}}}{\frac{{\mu_{I}^{(2)}/I^{(2)}}%
}{{\mu_{I}^{(1)}/I^{(1)}}}}\approx1+{\epsilon_{BW}^{(1)}}-{\epsilon_{BW}%
^{(2)}}}%
\end{equation}
where $\mu_{I}$ is the nuclear magnetic dipole moment, and I the nuclear spin
for the isotopes involved. The experimental $a$-constants should be corrected for second order hyperfine interaction. However, the value of the $a$-constants is fairly insensitive to this correction, so the only cases when this will have an effect is then the correction is large, that is when the experimental error is large, due to large errors in the fitting of the $a$-constant.

Using the effective operator formalism \cite{lindgren,sandars}, the hyperfine
interaction is divided into three parts, orbital, spin-dipole and contact
(spin) interaction. The hyperfine interaction constants can then be expressed
as a linear combination of effective radial parameters $a^{ij}_{l}$ and
angular coefficients $k^{ij}_{l}$,
\begin{equation}
a(J)=k^{01}_{l}a^{01}_{l}+k^{12}_{l}a^{12}_{l}+k^{10}_{l}a^{10}_{l}+k^{10}%
_{s}a^{10}_{s}%
\end{equation}
where the indices stand for the rank in the spin and orbital spaces,
respectively. Of these, only the contact interactions (10) of s and $p_{1/2}$
electrons contribute to the hyperfine anomaly.

We can thus rewrite the general magnetic dipole hyperfine interaction constant
in a simpler form when dealing with hyperfine anomaly;
\begin{equation}
a=a_{nc}+a_{s}+a_{p}=a_{nc}+k^{10}_{s}a^{10}_{s}+k^{10}_{p}a^{10}_{p}%
\end{equation}
where $a_{s}$ and $a_{p}$ are the contributions due to the contact interaction
of $s$ and $p_{1/2}$ electrons, respectively, and $a_{nc}$ is the contribution
due to non-contact interactions. The experimentally determined hyperfine
anomaly, which is defined with the total magnetic dipole hyperfine interaction
constant $a$, should then be rewritten to obtain the relative contributions to
the hyperfine anomaly:
\begin{equation}
{^{1}\Delta^{2}_{exp}}={^{1}\Delta^{2}_{s}}{\frac{a_{s}}{a}+^{1}\Delta^{2}_{p}}{\frac{a_{p}}{a} }%
\end{equation}

where ${^{1}\Delta^{2}_{s}}$ and ${^{1}\Delta^{2}_{p}}$ are the hyperfine
anomaly for an s- and p-electrons, respectively.

By performing an analysis of the hyperfine interaction it is possible to
deduce the different contributions to the hyperfine interaction constants, and
thus the hyperfine anomaly. That is the experimental hyperfine anomaly, which
might show a J-dependence, can be used to extract the hyperfine anomaly for an
s- (or p-) electron, ${^{1}\Delta^{2}_{s}}$.

It has been shown by Persson \cite{persson}, that it is possible to extract
the hyperfine anomaly without knowing the nuclear magnetic dipole moments,
provided you know the contribution of the contact interaction to the hyperfine
interaction constant in two atomic states;%

\begin{equation}
\label{eqn}{\frac{{a^{(1)}_{A}/a^{(2)}_{A}}}{{a^{(1)}_{B}/ a^{(2)}_{B}}}}
\approx{1+{^{1}\Delta^{2}_{s}} ({{\frac{a_{s}^{A}}{a^{A}}} -{\frac{a_{s}^{B}%
}{{a^{B}}}})}}%
\end{equation}

where A and B are two atomic states in the isotopes 1 and 2. The original use
was for radioactive isotopes where the atomic factor, (${{\frac{a_{s}^{A}%
}{a^{A}}}-{\frac{a_{s}^{B}}{{a^{B}}}}}$), were calibrated to a known hyperfine
anomaly between two stable isotopes. However, it is possible to use this
method on stable isotopes where the nuclear magnetic dipole moment is not
known with high accuracy. I will show this by applying the method to Gd and Nd.

\section{Hyperfine structure}

\label{sec:1}

Over the years many investigations of the hyperfine structure have been
carried out in the rare-earth (4f-shell) region \cite{pfeufer}, and a vast
amount of hyperfine interaction constants and isotope shift data has been
obtained. One would expect that analysis of the hyperfine interaction in this
region is difficult due to the large number of states. However, one finds that
many states are very close to pure LS-coupling, especially the low-lying
states. It is therefore relatively easy to perform an analysis. Even if the
hyperfine structure has been well studied, the nuclear magnetic dipole moments
are not always known to high accuracy. The nuclei in this region are often
deformed leading to a large nuclear electric quadrupole moment and drastical
changes in mean charge square radius \cite{otten}. It is therefore interesting to study the hyperfine anomaly, both in stable and unstable isotopes.

\subsection{Hyperfine structure in Nd}

\label{sec:Nd}

The hyperfine structure of Nd has been studied with high accuracy by Childs
and Goodman \cite{CG72} and Childs \cite{childs}, in the $4f^{4}6s^{2}%
\;^{5}\!I_{4-8}$ and $4f^{4}5d6s\;^{7}\!L_{5-11},^{7}\!K_{4}$ states with the
ABMR and LRDR methods, respectively. The studied states have also been found
to be close to LS-coupling (98-100 \% pure), making an analysis of the hyperfine interaction
rather simple. The states show no sign of large second-order hyperfine interaction, since the experimental errors for the $a$- \& $b$-constants are small. The high accuracy of the hyperfine interaction constants and
that an analysis has been performed \cite{childs} makes it possible to use
equation \ref{eqn}. It is important that the atomic factor (${{\frac{a_{s}%
^{A}}{a^{A}}}-{\frac{a_{s}^{B}}{{a^{B}}}}}$) attains a relatively large value,
in order to avoid errors \cite{persson}. It is therefore important to choose
the atomic states in the analysis with great care. From the the ratios of the
$a$ constants between the two stable isotopes $^{143}\!Nd$ and $^{145}\!Nd$,
we choose the $4f^{4}5d6s\;^{7}\!L_{5}$ and $4f^{4}5d6s\;^{7}\!K_{4}$ states
for extraction of the hyperfine anomaly as states B in equation \ref{eqn}, and use
$4f^{4}5d6s\;^{7}\!L_{11}$ as state A. The experimental $a$-constants with ratios and
contact contribution are given in table \ref{tab:2}. The hyperfine anomaly for
s-electrons are deduced and the result presented in table \ref{tab:5}. The
error of the hyperfine anomaly is only due to the experimental errors of the
hyperfine interaction constants, as the errors in the contact contribution (e.i. the eigenfunctions) are
not known. The error of the hyperfine anomaly is therefore too small and
should be about two to four times larger when the uncertainties of the contact
contributions are known. Once the hyperfine anomaly is determined it is
possible to use this result as a way of obtaining the ratio of the nuclear
magnetic dipole moments. The nuclear magnetic dipole moment has been measured
by Smith and Unsworth \cite{nn} using the ABMR technique and the ratio is given with the calculated
ratio in table \ref{tab:5}. In addition, Halford \cite{halford} has obtained the ratio of the nuclear magnetic moments using the ENDOR technique, which is also given in table \ref{tab:5}. We also note that the $a$-constants ratio in
$4f^{4}6s^{2}\;^{5}\!I_{4-8}$ states, is the same as the "new" ratio, this is
an indication that the hyperfine anomalies in these states are zero. This is
not surprising, as the configuration does not contain an unpaired s-electron,
thus having no contact interaction. This has also been found to be the case in
other rare-earths. The agreement between the ENDOR measurements and the value obtained here is good taking the errors into account, showing the validity of the method. Using these values it is possible to determine the hyperfine anomaly in other Nd isotopes as outlined by Anjun et al.\cite{Anjun}.
\begin{table}[ptb]
\caption{Hyperfine interaction constants in Nd.}%
\label{tab:2}
\centering
\begin{tabular}
[c]{lccccc}\hline
& $a \;^{143}\!Nd (MHz) $ & $a\; ^{145}\!Nd (MHz)$ & $a \;^{143}\!Nd/ a\;
^{145}\!Nd$ & $k_{s}$ & ${a_{s}/a}$\\\hline
$4f^{4}5d6s \;^{7}\!L_{11}$ & -151.318(1) & -93.982(1) & 1.610074(20) &
0.0454545 & 0.45176\\
$4f^{4}5d6s \;^{7}\!L_{5}$ & -55.216(1) & -34.482(1) & 1.601299(55) &
-0.0833333 & -2.26981\\
$4f^{4}5d6s \;^{7}\!K_{4}$ & -46.805(1) & -29.289(1) & 1.598040(65) & -0.1 &
-3.21333\\\hline
\end{tabular}
\end{table}

\begin{table}[ptb]
\caption{Hyperfine anomaly and magnetic dipole ratios in Nd.}%
\label{tab:5}
\centering
\begin{tabular}
[c]{lccccc}\hline
& ${^{143}\Delta^{145}_{s}}(\%)$ & $\mu_{I}\;^{143}\!Nd /\mu_{I}\;^{145}\!Nd$
&  &  & \\\hline
$4f^{4}5d6s \;^{7}\!L_{11}$ &  & 1.60860(3) &  &  & \\
$4f^{4}5d6s \;^{7}\!L_{5}$ & 0.2013(57) & 1.60871(6) &  &  & \\
$4f^{4}5d6s \;^{7}\!K_{4}$ & 0.2055(68) & 1.60853(7) &  &  & \\
$mean$ & 0.2034(63) & 1.60861(6) &  &  & \\
$ABMR\cite{nn}$ &  & 1.626(12) &  &  & \\
$ENDOR\cite{halford}$ & & 1.60883(4) & & &\\
$4f^{4}6s ^{2} \;^{5}\!I (mean)\cite{CG72}$ &  & 1.60861(2) &  &  & \\\hline
\end{tabular}
\end{table}

\subsection{Hyperfine anomaly in Gd}

\label{sec:Gd}

Precise studies of the hyperfine structure in Gd have been performed by
Unsworth \cite{uns}, who measured the $^{9}D$ term in the $4f^{7}5d6s^{2}$
configuration, and Childs \cite{childs1}, who studied the $4f^{7}%
5d^{2}6s\;^{11}\!F$ term. Unsworth \cite{uns} found that the contact
interaction was very small, indicating that no core-polarisation is present in
the $4f^{7}5d6s^{2}$ configuration. The same have also been found in other
rare-earths and this seem to be a general feature.
In addition should the
$4f^{7}5d6s^{2}\;^{9}D$ states exhibit no hyperfine anomaly, something that
can also be seen from the lack of J-dependence of the $a$ constant ratios for
the two isotopes $^{155,157}\!Gd$. The levels $4f^{7}5d^{2}6s\;^{11}\!F$ term
is reported to be 98-99\% pure L-S coupled states \cite{NBS}, making it
possible to use pure L-S coupling in the analysis of the hyperfine interaction
\cite{childs1}. In table \ref{tab:3} the $a$ constants for the $^{11}%
\!F_{2,3,8}$ states are given together with the s-electron contact
contribution. Using equation \ref{eqn} it is possible to derive the hyperfine
anomaly, the result is given in table \ref{tab:4}. The ratio of the nuclear
magnetic dipole moments has also been extracted. The ratio obtained compares
well with the ratio of the $4f^{7}5d6s^{2}\;^{9}\!D$ states, proving that the
ratio is close to the actual ratio and that there is no hyperfine anomaly in
these states. The ratio is also in agreement with the ENDOR measurement by Baker et al.
\cite{baker}, but the experimental error is one order of magnitude larger than
the derived error. Baker et al. \cite{baker} were also able to deduce the hyperfine anomaly of $Gd^3+$ in $CeO_{3}$ crystals. They obtained a value of ${^{157} \Delta^{155}_{s}}=0.07(12){\%}$ in agreement with the present result.
\begin{table}[ptb]
\caption{Hyperfine interaction constants in Gd.}%
\label{tab:3}
\centering
\begin{tabular}
[c]{lccccc}\hline
& $a \;^{157}\!Gd (MHz) $ & $a\; ^{155}\!Gd (MHz)$ & $a \;^{157}\!Gd/ a\;
^{155}\!Gd$ & $k_{s}$ & ${a_{s}/a}$\\\hline
$4f^{7}5d^{2} 6s \;^{11}\!F_{2}$ & -227.108(2) & -172.942(2) & 1.31320(5) &
0.4 & 1.2294(36)\\
$4f^{7}5d^{2} 6s \;^{11}\!F_{3}$ & -161.933(2) & -123.333(2) & 1.31297(5) &
0.25 & 1.0776(32)\\
$4f^{7}5d^{2} 6s \;^{11}\!F_{8}$ & -106.124(2) & -80.849(2) & 1.31262(5) &
0.125 & 0.8222(25)\\\hline
\end{tabular}
\end{table}

\begin{table}[ptb]
\caption{Hyperfine anomaly and magnetic dipole ratios in Gd.}%
\label{tab:4}
\centering
\begin{tabular}
[c]{lccccc}\hline
& ${^{157} \Delta^{155}_{s}}(\%)$ & $\mu_{I}\;^{155}\!Gd /\mu_{I}\;^{157}\!Gd$
&  &  & \\\hline
$4f^{7}5d^{2} 6s \;^{11}\!F_{2}$ & 0.108(18) & 0.76249(24) &  &  & \\
$4f^{7}5d^{2} 6s \;^{11}\!F_{3}$ & 0.104(28) & 0.76250(24) &  &  & \\
$4f^{7}5d^{2} 6s \;^{11}\!F_{8}$ &  & 0.76250(24) &  &  & \\
$mean$ & 0.106(24) & 0.76250(24) &  &  & \\
$ENDOR \cite{baker}$ &  & 0.7633(45) &  &  & \\
$4f^{7}5d6s ^{2} \;^{9}\!D (mean)\cite{uns}$ &  & 0.76254(40) &  &  & \\\hline
\end{tabular}
\end{table}

\section{Conclusions}

\label{con}

The method of deriving hyperfine anomaly with out knowing the nuclear magnetic moments \cite{persson} has been applied
to $^{155,157}$Gd and $^{143,145}$Nd, giving the first precise values of the hyperfine
anomaly for these isotopes. In addition have more precise values of the
nuclear magnetic dipole moment ratios been obtained. These values are in
agreement with the $a$-constant ratios for the ground terms $4f^{7}5d6s ^{2}
\;^{9}\!D$(Gd) and $4f^{4}6s ^{2} \;^{5}\!I$(Nd), proving that the hyperfine
anomalies in these states are negligible and that the electron core
polarisation in the ground terms is very small. Using this method makes it
possible to obtain values of the nuclear magnetic dipole moment ratio as well
as a value of the hyperfine anomaly solely from hyperfine interaction constants. This is of special interest in the case of
systematic studies over long chains of isotopes, to obtain information of
nuclear structure.

\end{document}